\documentclass[pra,twocolumn,a4paper,showpacs,superscriptaddress]{revtex4}
\usepackage{amsmath}
\usepackage{amsfonts}
\usepackage{graphicx}
\usepackage{longtable}
\begin{document}

\title{Equivalence of classicality and separability based on P phase-space representation of symmetric multiqubit states}  
\author{A. R. Usha Devi}
\email{arutth@rediffmail.com}
\affiliation{Department of Physics, Bangalore University, 
Bangalore-560 056, India}
\affiliation{Inspire Institute Inc., Alexandria, Virginia, 22303, USA.}
\author{A. K. Rajagopal} 
\affiliation{Harish-Chandra Research Institute, Chhatnag Road, Jhunsi, Allahabad 211 019, India}
\affiliation{Inspire Institute Inc., Alexandria, Virginia, 22303, USA.}
\author{Sudha}
\affiliation{Inspire Institute Inc., Alexandria, Virginia, 22303, USA.}
\affiliation{Department of Physics, Kuvempu University, Shankaraghatta, Shimoga-577 451, India.}
\author{H. S. Karthik}
\affiliation{Raman Research Institute, Bangalore 560 080, India}
\author{J. Prabhu Tej} 
\affiliation{Department of Physics, Bangalore University, 
Bangalore-560 056, India}

\date{\today}
\begin{abstract} 
Classical and quantum world views differ in peculiar ways. Understanding decisive quantum features -- for which no classical explanation exist -- and their interrelations is of foundational interest. Moreover, recognizing non-classical features carries practical significance in information processing tasks as it offers insights as to why quantum protocols work better than their classical counterparts.  We focus here on  two celebrated notions of non-classicality  viz.,  negativity of P phase-space representation and entanglement in symmetric multiqubit systems. We prove that they imply each other.  
\end{abstract}
\maketitle  

\section{Introduction}
Contrast between quantum and classical conceptions of nature remains to be an enigmatic topic from the early days of quantum theory. Identifying hall-mark demarcations between the two descriptions continues to draw intensive attention. Gaining insights into the distinct features of the two theories attain fundamental significance in quantum information science as it is crucial to know why/how several quantum information protocols  outperform the  corresponding classical ones.

Departure of the  P phase space representation~\cite{gsp} from being a true classical probability density function  offers a decisive signature of non-classicality in single mode bosonic systems. Yet another striking aspect projected out by the quantum world view is non-classical correlations in composite systems. It is important to explore if these two illustrious notions of non-classicality are interwoven with each other. However, it has been shown recently that the two concepts disagree maximally in bosonic systems~\cite{paris}.  In this paper we prove that  classicality implied by positive P representation for spins~\cite{Arecchi} and separability~\cite{werner} imply each other in  symmetric multiqubit systems. 

P representation  allows a decomposition of the spin density matrix as a weighted sum of spin coherent states (or SU(2) coherent states)~\cite{Arecchi}.  Classicality of spin states, based on well behaved P representation (i.e., P function obeying the properties of a true probability density) has been explored by Giraud et. al.~\cite{braun}. Luis and Rivas~\cite{Rivas}  showed that spin squeezing~\cite{ueda} manifestly reflects the failure of P function to be a classical probability density. Simple operational procedures revealing the violation of classical statistical bounds by spin states endowed with non-classical P representation were also derived in Ref.~\cite{Rivas}. Besides studying the non-classicality of single spin systems, Giraud et. al.~\cite{braun} also explored the implications of P representation on quantum correlations in bipartite spin systems. They showed that a composite system consisting of two qubits is separable if and only if their P function is positive. However, in the case of a coupled system consisting of spin-1/2 and spin-1, it was shown~\cite{braun} that separable states can exhibit non-classicality (non-positive or singular P function). Here, we confine our attention to permutation symmetric multiqubit states and show that  non-classicality implies entanglement and vice versa.    
  
\section{ The P representation of quantum electromagnetic field}

  In 1963, Sudarshan and Glauber introduced the celebrated diagonal coherent state representation or the P representation~\cite{gsp} to characterize states of quantized radiation fields in terms of a classical phase-space like description. In this representation the density matrix $\hat\rho$ is expressed in terms of a classical function $P(\alpha)$     
\begin{equation} 
\label{P}
\hat\rho=\int{\rm d}^2 \alpha\, P(\alpha)\, |\alpha\rangle\langle \alpha|.
\end{equation} 
where $|\alpha\rangle=\hat D(\alpha)\vert 0\rangle=e^{-\vert\alpha\vert^2/2}\, \displaystyle\sum_{n=0}^{\infty}\, \frac{\alpha^n}{\sqrt{n!}}\vert n\rangle$, are coherent states of light; $\hat D(\alpha)=e^{\hat a^\dag\alpha-\hat a\alpha}$ denotes the displacement operator and $\vert 0\rangle$ is the vacuum state of the radiation field.  

The P phase-space representation  reproduces  statistical averages of normally ordered operators in a classically equivalent manner. However, P-function cannot be interpreted as classical probability density as it can assume negative values or is more singular than delta function  for {\em non-classical} radiation states. This feature brings forth the inevitability of a fully quantum description. 
   
For coherent states of radiation $\hat\rho=\vert \alpha' \rangle\langle\alpha'\vert$, the P function is a delta function i.e., $P(\alpha)=\delta(\alpha-\alpha')$. This leads to the interpretation that coherent states allow for  as classical a description as possible. 
The P-function of convex mixtures of coherent states $\hat{\rho}=\sum_i\, p_i\ |\alpha _i\rangle\langle \alpha _i |,\ \ 0\leq p_i\leq 1, \sum_i p_i=1$ is given by  $P(\alpha)=\sum_i\  p_i\, \delta(\alpha-\alpha _i)$ and such a mixture is also classical by the same token. The  P representation of  radiation has led to  well corroborated notions of non-classicality.        

\section{Spin coherent state representation} 

Along similar lines of the phase space distribution functions for radiation states,  Arecchi et. al.~\cite{Arecchi} introduced an analog of diagonal spin coherent state representation:  
\begin{equation} 
\label{pdef}
\hat\rho=\int{\rm d}\Omega\, P(\theta,\phi)\, |\theta,\phi\rangle\, \langle \theta,\phi|,
\end{equation}
where ${\rm d}\Omega=\sin\theta{\rm d}\theta\, {\rm d}\phi$;  $|\theta,\phi\rangle$ denote the spin coherent states (SCS)  
\begin{equation}
\label{scs}
|\theta,\phi\rangle={\rm exp}(\tau\hat S_+ - \tau^*\, S_-)\, \vert S, -S\rangle, \ \tau=\frac{1}{2}\, \theta\  e^{-i\phi}.
\end{equation} 
 Here $\hat S_\pm=\hat S_x\pm i \hat S_y$ are ladder spin operators; $\{\vert S,\, M\rangle, -S\leq M\leq S\}$ denote simultaneous eigen states of the squared spin operator $\hat{S}^2$ and $z$ component of spin $\hat{S}_z$.   
 
Identifying~\cite{Var} that  ${\rm exp}~(\tau\hat S_+ - \tau^*\, S_-)=\hat R(\phi-\pi, \theta, \pi-\phi)=
e^{-(\phi-\pi)\, \hat S_z}\, e^{-\theta\, \hat S_y}\, e^{-(\pi-\phi)\, \hat S_z}$,    with    
$(\phi-\pi, \theta, \pi-\phi)$ denoting the Euler angles of rotation,  the spin coherent states can be expressed as    
\begin{widetext}
\begin{eqnarray}
\label{scs1}
|\theta,\phi\rangle&=& \hat R(\phi-\pi, \theta, \pi-\phi)\, \vert S, -S\rangle \nonumber \\ 
&=& \displaystyle\sum_{M=-S}^{S}\, D^S_{M, -S}(\phi-\pi, \theta, \pi-\phi)\, \vert S, M\rangle \nonumber \\ 
\label{scs2}
&=& \displaystyle\sum_{M=-S}^{S}\, \sqrt{\left(\begin{array}{c} 2S \\ S+M\end{array}\right)} 
\left( \cos\frac{\theta}{2}\right)^{S-M}\, \left(\sin\frac{\theta}{2}\right)^{S+M}\, e^{-i(S+M)\phi}\vert S, M\rangle, 
\end{eqnarray} 
\end{widetext}
where $D^S_{M, -S}(\phi-\pi, \theta, \pi-\phi)=\langle S, M\vert \hat R(\phi-\pi, \theta, \pi-\phi)\vert 
S, -S\rangle$ denotes the Wigner D-function~\cite{Var}.  

The unit trace condition ${\rm Tr}[\hat\rho]=1$ leads to (see Eq.~(\ref{pdef}))   
\begin{equation}
\int d\Omega\, P(\theta, \phi)=1,
\end{equation}
as spin coherent states are normalized: $\langle \theta,\phi\vert \theta, \phi\rangle=1.$
However, the P function is not ensured to be a bonafide statistical probability density -- and hence it leads to notions of non-classicality of spin states~\cite{braun, Rivas}.

Any arbitrary pure state of a qubit (spin-1/2 system) $\vert \psi\rangle = \cos(\Theta/2)\, e^{-i\Phi}\vert 0\rangle + \sin(\Theta/2)\, e^{i\Phi}\vert 1\rangle$ is a spin coherent state as it can always be expressed as a rotated "spin down state" $\vert \psi\rangle=\hat {\cal R}(\pi-\Phi, \pi+\Theta, \Phi-\pi)\vert 1\rangle\equiv\vert \pi+\Theta, -\Phi\rangle.$ And the P function of a pure qubit state is readily identified to be,  
$P_\psi(\theta, \phi)=\delta(\theta-\theta')\delta(\phi-\phi')$;\  $\theta'=\pi+\Theta,\ \phi'=-\Phi$. In other words, pure state of a qubit is {\em classical} because its P function is positive. 

Further, any mixed state of a qubit can be expressed in terms of its spectral decomposition  
\begin{eqnarray} 
\hat\rho&=& p \, \vert \psi\rangle\langle \psi\vert + (1-p)\, \vert \psi_\perp\rangle\langle \psi_\perp\vert \nonumber \\
&=& p\, \vert \theta', \phi'\rangle\langle \theta', \phi'\vert  + (1-p) \vert \theta'', \phi''\rangle\langle \theta'', \phi''\vert       
\end{eqnarray}  
which brings forth a convex decomposition of the density matrix in terms of spin coherent states. Corresponding P function is easily identified to be 
\begin{equation}
P(\theta, \phi)\equiv p \, \delta(\theta-\theta')\delta(\phi-\phi') + (1-p)\, \delta(\theta-\theta'')\delta(\phi-\phi'')  
\end{equation}
with $0\leq p\leq 1$. In other words, a qubit is {\em classical} as it is always P representable~\cite{braun}. This observation indicates that non-classicality in a system of qubits could be essentially attributed to quantum correlations amongst qubits.  We proceed further to prove that this is indeed the case, when symmetric multiqubit systems are concerned.
   
\section {Non-classicality of symmetric $N$ qubit systems}
  
For a system of $N$ qubits, the collective spin operators  $\hat S_\mu$, $(\mu=x,\,y,\,z)$ are expressed in terms of $N=2\, S$ Pauli spin operators $\frac{1}{2}\, \sigma_{i\mu}$,
\begin{equation} 
\hat S_\mu= \frac{1}{2}\sum_{i=1}^{N}\,\sigma_{i\mu}, \ \ \ \mu=x,\,y,\,z 
\end{equation} 
and the collective  angular momentum eigenstates  $\{\vert S=N/2, M\rangle,\ -S\leq M\leq S\}$, the common eigenstates of $\hat {S}^2$ and $\hat{S}_z$  correspond to the {\em maximal value of total angular momentum} resulting in the addition of $N$ spin-1/2 constituents. This $(2S+1)=N+1$ dimensional subspace of $2^N$ dimensional Hilbert space of $N$ spin-1/2 particles (qubits) corresponds to {\em permutation symmetric states} of $N$ qubits. Any arbitrary state 
$\vert \Psi\rangle=\sum_{M=-S}^{S}\, C_M\, \vert S=N/2, M\rangle$ in this subspace is a symmetric multiqubit state. 

In terms of the constituent qubit states, the spin coherent state is expressed as, 
\begin{widetext}
\begin{eqnarray}
\label{scsmq}
|\theta,\phi\rangle&=& \hat R(\phi-\pi, \theta, \pi-\phi)\, \vert S, -S\rangle \nonumber \\
&=&  \hat {\cal R}(\phi-\pi, \theta, \pi-\phi)\otimes \hat {\cal R}(\phi-\pi, \theta, \pi-\phi)\otimes\cdots\, \otimes\hat {\cal R}(\phi-\pi, \theta, \pi-\phi) \vert 1, 1,\cdots , 1\rangle\nonumber \\ 
&=&\vert 1_{(\theta,\phi)},\,  1_{(\theta,\phi)},\cdots  1_{(\theta,\phi)}\rangle   
\end{eqnarray} 
\end{widetext}
where $\hat {\cal R}(\phi-\pi, \theta, \pi-\phi)$ is the rotation operator on individual qubits and $\vert 1_{(\theta,\phi)}\rangle=\hat {\cal R}(\phi-\pi, \theta, \pi-\phi)\vert 1\rangle$ is a rotated "spin-down" qubit state. 
 
The P-representation of symmetric $N$-qubit states thus assumes the form,  
\begin{eqnarray}
\label{PMQ}
 \hat{\rho}_{\mbox{sym}}(N)&=&\int{\rm d}\Omega\, P(\theta,\phi)\, |\theta,\phi\rangle\, \langle \theta,\phi| \nonumber \\ 
 &=& \int{\rm d}\Omega\, P(\theta,\phi)\, \hat\rho_{(\theta,\phi)}\otimes \hat\rho_{(\theta,\phi)}\otimes 
 \ldots \otimes \hat\rho_{(\theta,\phi)} \nonumber \\
\end{eqnarray} 
where $\hat\rho_{(\theta,\phi)}=\vert 1_{(\theta,\phi)}\rangle\langle 1_{(\theta,\phi)}\vert$. It is clear that if  $P(\theta, \phi)$   is positive, the symmetric multiqubit state $\hat{\rho}_{\mbox{sym}}$  admits a  separable decomposition. Conversely, we show that the set of {\em all} separable symmetric multiqubit states  exhibit a positive P representation. 

Separable symmetric $N$-qubit states have the form, 
\begin{equation} 
\label{sepsymN}
\hat\rho^{\rm (sep)}_{\rm sym}(N)=\sum_{w}\ p_w\, \hat\rho_w\otimes \hat\rho_w\otimes \ldots \otimes \hat\rho_w 
\end{equation} 
where $\sum_i p_w=1, 0\leq p_w\leq 1$ and $\rho_w$ denotes the constituent single qubit density matrices:
\begin{equation}
\rho_w=\frac{1}{2}[I + \sum_{\mu=x,y,z}\, \sigma_\mu\, s_{w\mu}]. 
\end{equation}
Positivity of the density matrix $\rho_w$ requires that the real parameters $s_{w\mu}={\rm Tr}[\hat\rho_w\, \sigma_\mu]$ obey the condition 
$\sum_\mu\, s^2_{w\mu}=s^2_{wx}+ s^2_{wy}+ s^2_{wz}\leq 1$ --  equality holds when the density matrix is pure. We continue now to prove that the constituent qubit density matrices $\hat\rho_w$ are all pure.  

We consider a two qubit reduced system of a symmetric separable state $\hat\rho^{\rm (sep)}_{\rm sym}(N)$, by tracing over all the other qubits: 
\begin{equation} 
\label{tqsep}
\varrho^{\rm (sep)}_{\rm sym}(2)=\sum_{w}\ p_w\, \hat\rho_w\otimes \hat\rho_w.
\end{equation}     
Comparing (\ref{tqsep}) with the general structure of an arbitrary symmetric two qubit density matrix~\cite{ARU}
\begin{widetext}
\begin{eqnarray}
\label{tq}
\rho_{\rm sym}(2)&=& \frac{1}{4}\left[I\otimes I+\sum_{\mu=x,y,z}\, s_{\mu} \, \sigma_\mu\otimes I+\sum_{\mu=x,y,z}\,s_{\mu}(I\otimes \sigma_\mu\,) +\sum_{\mu,\nu=x,y,z}\, t_{\mu\,\nu}\,  \sigma_\mu\otimes \sigma_\nu\, \right]   
\end{eqnarray}
\end{widetext}
where 
\begin{eqnarray*}
s_{\mu}&=&{\rm Tr}[\rho_{\rm sym}(2)\, \sigma_\mu\otimes I]={\rm Tr}[\rho_{\rm sym}(2)\, I\otimes \sigma_\mu] \\
 t_{\mu\nu}&=&{\rm Tr}[\rho_{\rm sym}(2)\, \sigma_\mu\otimes \sigma_\nu]={\rm Tr}[\rho_{\rm sym}(2)\, \sigma_\nu\otimes \sigma_\mu]=t_{\nu\mu} 
\end{eqnarray*}
are the real parameters characterizing the two qubit symmetric density matrix. The correlation parameters $t_{\mu\nu}$ of a symmetric two qubit state obey the following condition~\cite{ARU}: 
\begin{equation}
 \label{trtcond}
  \sum_\mu t_{\mu\mu}=1,
 \end{equation} 
 and this condition plays a crucial role in identifying that the constituent qubit density matrices $\hat\rho_w$ of a separable symmetric state (\ref{sepsymN}) are pure. 
 
It is easy to find that in a separable symmetric two qubit state (\ref{tqsep}), the  parameters $s_\mu$ and $t_{\mu\nu}$ are given by, 
\begin{equation}
s_\mu=\sum_{w}\, p_w\, s_{w\mu}, \ \ t_{\mu\nu}=\sum_{w}\, p_w\, s_{w\mu}\, s_{w\nu} 
\end{equation} 
and the condition (\ref{trtcond}) implies that 
\begin{equation}
\sum_\mu t_{\mu\mu}=\sum_{w}\, p_w\, \left\{\sum_\mu\, s^2_{w\mu}\right\}=1\Rightarrow \sum_\mu\, s^2_{w\mu}=1. 
\end{equation}  
In other words, the single qubit density matrices $\hat\rho_w$ of a separable symmetric multiqubit system are pure. This, in turn, would lead us to identify that $\hat\rho_w=\vert 1_{(\theta_w,\phi_w)} \rangle \langle 1_{(\theta_w,\phi_w)}\vert
$ as all single qubit pure states are spin coherent states themselves. Thus, we identify that separable $N$ qubit symmetric states admit a convex decomposition in terms of spin coherent states:    
\begin{eqnarray} 
\label{sepd}
\hat\rho^{\rm (sep)}_{\rm sym}(N)&=&\sum_{w}\ p_w\, \hat\rho_{(\theta_w,\phi_w)}\otimes \hat\rho_{(\theta_w,\phi_w)}\otimes \ldots \otimes \hat\rho_{(\theta_w,\phi_w)} \nonumber \\ 
&=& \sum_{w}\ p_w\, \vert \theta_w,\phi_w\rangle \langle \theta_w,\phi_w\vert.
\end{eqnarray} 
It is easy to recognize that the P function of separable symmetric multiqubit state is a well behaved convex sum:  
\begin{equation} 
P(\theta,\phi)=\sum_w\, p_w\, \delta(\theta_w-\theta)\, \delta(\phi_w-\phi). 
\end{equation} 

Entangled symmetric multiqubit states cannot be decomposed as mixtures of coherent states  with positive weights as in (\ref{sepd}), are   {\em non-classical}. More specifically, entanglement manifests itself in terms of  non-positive P function in symmetric multiqubit states.  

\section{Concluding remarks} 

Understanding non-classicality of a quantum system  has been of foundational significance from the early days of quantum theory and it has gained  rekindled interest in recent years owing to path-breaking implications in quantum information science. Clear conceptual demarcations that reflect the inevitablity of  quantum description -- with no classical analogue -- gains utmost significance from this point of view.  Two celebrated notions that address the issue are non-classicality based on P phase space representation and  non-classical correlations in composite systems discerned via established concepts from quantum information theory. A comparison of these two approaches in bosonic systems revealed maximal inequivalence in continuous variable systems~\cite{paris}. This motivated our investigation in symmetric multiqubit systems. We have identified here that the notion of non-classicality arising from a ill behaved P representation is synonymous with entanglement in symmetric multiqubits.


\begin{thebibliography}{0}
\bibitem{gsp} E. C. G. Sudarshan, \prl {\bf 10}, 277 (1963);  Roy J. Glauber, Phys. Rev. {\bf 131}, 2766 (1963). 
\bibitem{paris} A. Ferraro and Matteo G.A. Paris, \prl {\bf 108}, 260403 (2012) 
\bibitem{Arecchi} F. T. Arecchi, E. Courtens, R. Gilmore and H. Thomas, Phys. Rev. {\bf 6}, 2211 (1972).
 \bibitem{werner} R. F. Werner, \pra {\bf 40}, 4277 (1989).
\bibitem{braun} O. Giraud, P. Braun and D. Braun, \pra {\bf 78}, 042112 (2008); New Journal of Physics {\bf 12}, 063005 (2010).    
\bibitem{Rivas} A. Luis, A. Rivas, \pra {\bf 84}, 042111 (2011).
\bibitem{ueda} M. Kitagawa and M. Ueda, \pra {\bf 47}, 5138 (1993).
\bibitem{Var} D. A. Varshalovich, A. N. Moskalev and V. K. Khersonkii, {\em Quantum Theory of Angular Momentum}, 
(World Scientific, Singapore, 1988.)  
\bibitem{ARU} A. R. Usha Devi, M. S. Uma, R. Prabhu and A. K. Rajagopal, Phys. Lett. A, {\bf 364}, 203 (2007); A. R. Usha Devi, M. S. Uma, R. Prabhu and Sudha, Int. J. Mod. Phys. A {\bf 17}, 2267 (2002). 
%
\end{thebibliography}
\end{document}